\documentclass[10pt]{article}

\usepackage{amsmath}
\usepackage{amssymb}
\usepackage{graphicx}
\usepackage{multirow}
\usepackage{cite}

\usepackage{color}
\usepackage{url}

\usepackage[labelfont=bf,labelsep=period,justification=raggedright]{caption}

\makeatletter
\renewcommand{\@biblabel}[1]{\quad#1.}
\makeatother

\date{}

\begin{document}

\begin{flushleft}
{\Large
\textbf{Temporal effects in trend prediction: identifying the most popular nodes in the future}
}
\\
Yanbo Zhou$^{1}$
An Zeng$^{2}$
Wei-Hong Wang$^{1}$
\\
\bf{1} College of Computer Science and Technology, Zhejiang University of Technology, Hangzhou, P. R. China
\\
\bf{2} School of Systems Science, Beijing Normal University, Beijing, P. R. China
\\
$\ast$ E-mail: Corresponding wwh@zjut.edu.cn
\end{flushleft}

\section*{Abstract}
Prediction is an important problem in different science domains. In this paper, we focus on trend prediction in complex networks, i.e. to identify the most popular nodes in the future. Due to the preferential attachment mechanism in real systems, nodes' recent degree and cumulative degree have been successfully applied to design trend prediction methods. Here we took into account more detailed information about the network evolution and proposed a temporal-based predictor (TBP). The TBP predicts the future trend by the node strength in the weighted network with the link weight equal to its exponential aging. Three data sets with time information are used to test the performance of the new method. We find that TBP have high general accuracy in predicting the future most popular nodes. More importantly, it can identify many potential objects with low popularity in the past but high popularity in the future. The effect of the decay speed in the exponential aging on the results is discussed in detail.


\section*{Introduction}
The emergence of online social media and rich user-generated content bring the information overload problem. The online content has become increasingly abundant and immediately available, and users cannot go through every piece of information to find the high quality ones~\cite{simon}. These high quality source, in general, have formidable power to impact opinions, culture, and policy, as well as advertising profit. Thus they usually attract a lot of attention and become eventually popular. The rapid development of the Internet results in the availability of a huge amount of data with time information, making it possible to study the popularity dynamic of the online content. It was found that the popularity of various pieces of content on the Web, like news \cite{news}, Twitter \cite{asur2011trends}, blog posts \cite{blogs}, videos \cite{youtube,popularityyoutube}, posts in online discussion forums \cite{bbs} and product reviews \cite{product}, vary significantly on temporal scales. In this context, the early identification of the eventual popular content becomes an important problem \cite{lerman2010predict}. It cannot only improve the user experience as the prediction save their searching time for high quality contents among unpopular ones, but also bring commercial profit for the online vendor as it helps them better manage their inventory.

Although the preferential attachment (PA) is a success in explaining the power-law distribution that widely found in real systems \cite{BA}, it performs not satisfying enough when applied to predict the future popularity (or degree) of nodes. For example, it was found in the citation networks that some papers can attract significantly more citations than the prediction from PA \cite{Nopa,Nopaim}. The ability of a node to attract new links is found to decay exponentially with time, both in citation networks \cite{matusPRL} and information access \cite{infoAccess}. Moreover, temporal analysis of the popularity dynamics of online content in Wikipedia \cite{busrtM} and micro blog\cite{busrtWeibo} shows the burst pattern. An extensive study of how the content's popularity grows and fades over time in online media has been presented in ref. \cite{timeSeriesShap}. Besides the experimental study of the temporal dynamic, some possible mechanisms that may contribute to the experimental finding are proposed, such as relevance and time decay \cite{matusPRL}, random popularity shifts \cite{busrtM} and human dynamics \cite{youtube}. All these studies have shown that the high cumulative degree of nodes is not a guarantee for large degree increase in the future.

In this paper, we focus on the trend prediction in complex networks, i.e. to identify the most popular nodes in the future. In the literature, there are some existing methods for this problem especially for a certain application field \cite{asur2012,diggp,zenganTp,wang2013quantifying}. For instance, in the popular online service \emph{Digg.com}, the initial growth popularity has been used to predict its later popularity \cite{diggp}. Popularity-based predictor has been designed for trend prediction and its performance is shown to be further enhanced if the user social network is incorporated \cite{zenganTp}. In this paper, we introduce a temporal-based predictor (TBP) which takes advantage of the time decay effect found in many empirical works. The validation of the method is conducted in three time-stamped data sets. The results show that the prediction precision is remarkably higher than that of PA. In addition, the new method is especially effective in identifying the potential nodes with low popularity in the past but high popularity in the future.

\section*{Materials and Methods}

The system we considered in this paper can be modeled by the bipartite network which consists of a set of users $U$ and a set of objects $O$. We use Latin letters for users and Greek letters for objects to distinguish them. A bipartite network can be represented by an adjacency matrix $A$, where elements $A_{i\alpha}$ are equal to 1 if user $i$ has collected object $\alpha$ and 0 otherwise. We consider snapshots of these networks at different time stamps by taking into account only the links established before a given time $t$, and we use $A(t)$ to denote the adjacency matrix at time $t$. The number of objects collected by user $i$ and the number of users who collected object $\alpha$ at time $t$ ($i.e.$, user degree and object degree) are computed as $k_i(t)=\sum_{\alpha}A_{i\alpha}(t)$ and $k_\alpha(t)=\sum_{i}A_{i\alpha}(t)$, respectively.

The popularity increase of object $\alpha$ in future $T_F$ time steps (i.e. the future time window) is then
\begin{equation}
\label{eq.1}
\Delta{k_\alpha(t,T_F)}=k_\alpha(t+T_F)-k_\alpha(t).
\end{equation}
For a suitably chosen value of $T_F$, this quantity can measure the temporal interest in object $\alpha$. The main goal of trend prediction in this paper is to identify the most popular objects in the future. To this end, we define a testing time $t$ and a future time window of length $T_F$, and rank all objects according to their popularity increase $\Delta{k_\alpha(t,T_F)}$. This ranking is considered as the true ranking of popularity in the future. A generic predictor will make use of the information before $t$ and assign prediction scores {$s_\alpha$} to all objects. These scores will be mapped into a predicted ranking. In general, the higher overlap of the predicted ranking and the true ranking, the better the predictor is.

\subsection*{Popularity-based Predictor}

Preferential attachment is a well-known mechanism of network evolution which assumes that the probability a node to attract a new link is proportional to its cumulative degree. In trend prediction, this means that objects which are popular at time $t$ are expected to have better chances to attract new links from users. This implies that the cumulative degree of an object $k_\alpha(t)$ is a good predictor of its future popularity increase. Considering the decaying interest in objects, the prediction scores can be set as the recent popularity of objects. The prediction score of an object at time $t$ can be calculated by $\Delta{k}_\alpha(t,T_P)$ where $T_P$ is the length of the considered history. Recently, a popularity-based predictor (PBP) \cite{zenganTp} has been proposed to combine the predictor $k_\alpha(t)$ and $\Delta{k}_\alpha(t,T_P)$. PBP has a tunable parameter $\lambda \in [0,1]$ to make the new predictor change smoothly from $k_\alpha(t)$ to $\Delta{k}_\alpha(t,T_P)$. Mathematically, the prediction score of PBP is computed as
\begin{equation}
\label{eq.2}
s_\alpha(t,T_P)=(1-\lambda) k_\alpha(t)+\lambda \Delta{k}_\alpha(t,T_P)=k_\alpha(t)-\lambda k_\alpha(t-T_P).
\end{equation}
This predictor simplifies to the total popularity method when $\lambda = 0$ and to the recent popularity method when $\lambda = 1$.

\subsection*{Temporal-based predictor}
The popularity-based predictor in fact considers all the recent popularity of objects, but weakens the influence of links an object received before $t-T_P$. In the literature, it has been found that the interest toward individual objects vanishes exponentially with time in some case \cite{matusPRL,ZhuDecay}. Therefore, it is too arbitrary to simply divide the time into two segments in the popularity-based predictor, as it may lose the detailed temporal information of real networks.

To make better use of the temporal information for trend prediction, we proposed a temporal-based predictor (TBP) in this paper. In TBP, we consider the mechanism that the influence of a link exponentially decays with time. An aging function is accordingly introduced to calculate the prediction scores:
\begin{equation}
\label{eq.3}
s_\alpha(t)=\sum_{i}{A_{i\alpha}(t)e^{\gamma(T_{i\alpha}-t)}},
\end{equation}
where $T_{i\alpha}$ denotes the time at which user $i$ select object $\alpha$. $\gamma$ is a positive parameter which controls the decay speed. A larger $\gamma$ indicates a faster decay, and $\gamma=0$ corresponds to the cumulative popularity without any decay. TBP preserves all the detailed temporal information in the network. By adjusting $\gamma$, we can study the temporal effects in the prediction of the future popularity.

\section*{Data Description}

To test the performance of TBP, we use three distinct real data sets: MovieLens, Netflix and Facebook in this paper.
Movielens and Netflix data sets contain movie ratings, and Facebook data set contains users' wall post relationships. MovieLens is provided by GroupLens project at University of Minnesota (www.grouplens.org). We use their 10 million ratings data set. Each user in MovieLens data set has at least 20 ratings. Netflix is a huge data set released by the DVD rental company Netflix for its Netflix Prize (www.netflixprize.com). The original data has 480189 users, 17770 objects and 100480507 ratings. Since the original Movielens and Netflix data sets are large, we extracted a small subset from each of them by randomly choosing some users who have rated at least 20 movies and took all movies they had rated. For both Movielens and Netflix, the ratings are given on the integer scale from 1 to 5 (from worst to best). We here only consider the ratings higher than 2 as a link. The final data consists of 5000 users, 7533 movies, and 864581 links in Movielens and 4960 users, 16599 movies, and 1249058 links in Netflix. Facebook data set contains a list of all the wall posts from the Facebook New Orleans networks \cite{konect:2014:facebook-wosn-wall,viswanath09,konect}. A link from one user to another corresponds that the user post on another user's wall. As the Facebook network is a unipartite directed network, here we mapped it to a bipartite network with a set of users and a set of users' walls (objects). If a user has posted on a wall, there will be a link between the user and the wall. The original data has $42390$ users , $39986$ objects and $876993$ links. Since user may written on his own wall, we remove these links to eliminate self-influence. The final data consists of $40981$ users, $38143$ objects and $855542$ links. For all of these three data sets, the time is counted by days. The characteristics of these data sets are summarized in table 1. All these three data sets are available from the Koblenz Network Collection \cite{konect-network}, and they are all free to use even for commercial purposes.

\section*{Evaluation Metrics}

We apply three metrics to give quantitative measurements of the predictors' performance: AUC, precision and novelty.

As the main point of this paper is to predict which objects will be popular in the future, only the top part of the ranking list should be considered when evaluating the performance of the predictors. We thus use a standard measure in information filtering literature named AUC \cite{AUC}, which evaluates a ranking list by calculating the relative position of its top $n$ objects. We select the top $n$ objects in the real future as a group of benchmark objects, and denoted it as set $B$. The other objects are in the complement group of $B$, which denoted as $B'$. Then the AUC is calculated as
\begin{equation}
AUC=\frac{1}{|B||B'|}\sum_{\alpha \in B}\sum_{\beta \in B'}I(s_\alpha,s_\beta),
\end{equation}
where
\begin{equation}
I(s_\alpha,s_\beta)=
\begin{cases}
0,  \qquad\mathrm{if } \quad s_\alpha<s_\beta \\

0.5,  \quad\mathrm{if}\quad s_\alpha=s_\beta \\

1,  \qquad\mathrm{if } \quad s_\alpha>s_\beta

\end{cases}
.
\end{equation}

AUC equals to one when all benchmark objects are ranked higher than the other objects, while AUC=0.5 corresponds to
a completely random object ranking list.

Another evaluation matric is called precision. It is defined as the fraction of objects in the top $n$ places of
the estimated ranking that appear also in the top $n$ places of the true ranking \cite{Precision}. The precision of
the predictor is defined as $P_n=D_n/n$, where $D_n$ indicates the number of common objects in the top $n$ places of the predicted ranking and the true ranking. It lies in the range [0, 1], the higher the better.

It is often the case that objects popular in the future time window $(t,t+T_F]$ were already popular in the past. Successful prediction of those objects can contribute to precision $P_n$. However, prediction of these objects brings much less benefit to users than the prediction of genuinely "new entries", i.e. objects that were missing in top $n$ in the past but they appear there in the future time window. We label the true number of those objects as $E_n$ and the number of those successfully identified by the predicted ranking as $C_n$, respectively. The rate of correct prediction of these new entries is $Q_n = C_n/E_n$. This allows us to measure how well a predictor is able to identify the potential objects. Here, we name $Q_n$ as novelty.

\section*{Results}

To obtain the final evaluation of the predictors' performance, we average results over 10 randomly selected $t$ for each data sets. To make sure that there is enough history information, $t$ is set as at least one year later than the first record in each data set. As all the predictors we considered in this paper are based on objects' history, we only consider the objects with at lest one link before the testing date $t$.

\begin{figure}[h!]
\centering
\scalebox{0.7}[0.7]{\includegraphics{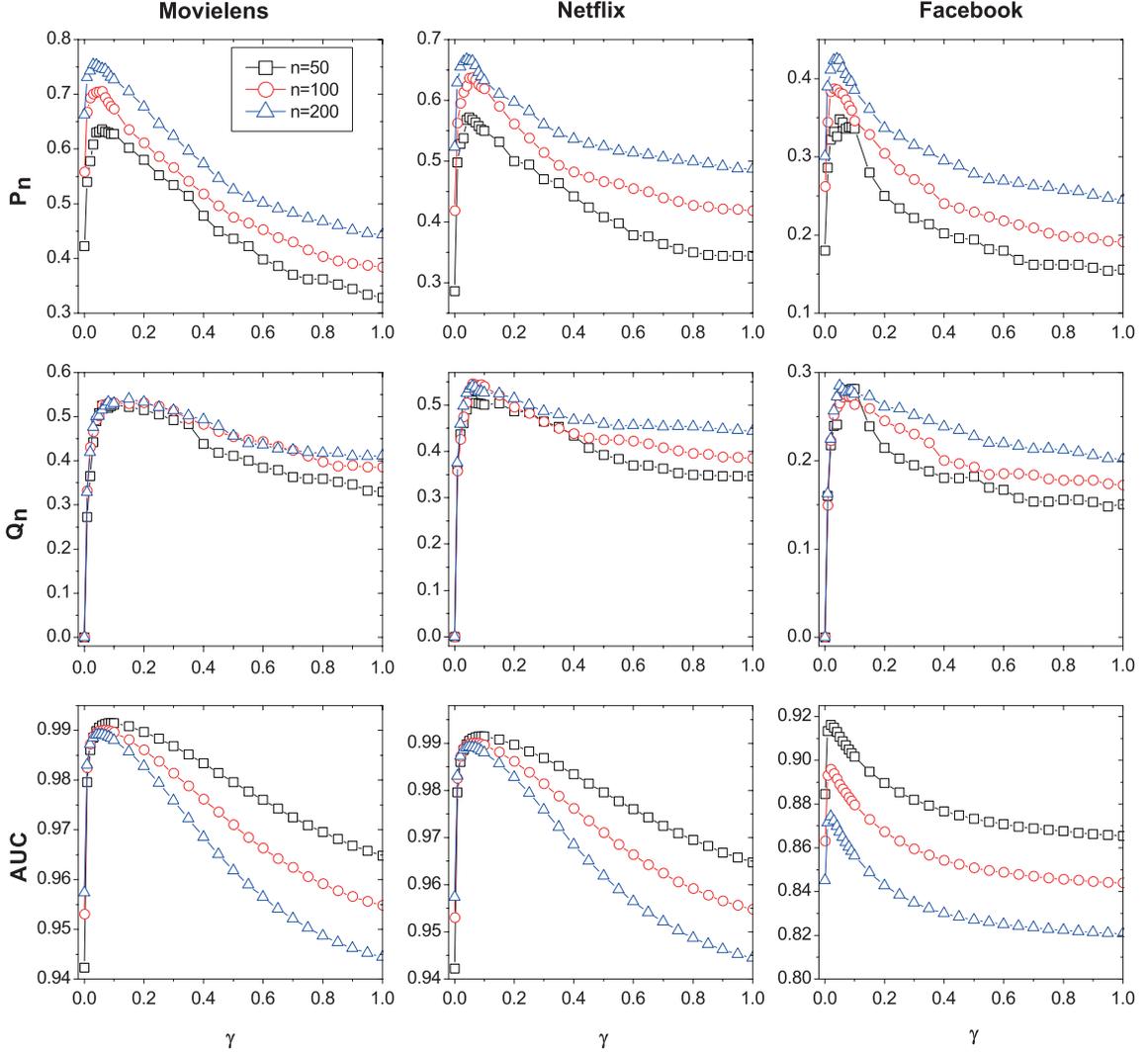} }
\caption{
{\bf The prediction result of TBP for Movielens, Netflix and Facebook data sets under different $\gamma$.} The performance of different $n$ values are given. $n=50$ is presented by black lines with squares, $n=100$ is presented by red lines with circles, and $n=200$ is presented by blue lines with triangles.}
\label{TBP}
\end{figure}

Figure \ref{TBP} shows the performance of the TBP under different $\gamma$ in Movielens, Netflix and Facebook data sets, respectively. $T_F$ is set as 30 days for all there data sets. Different $n$ values are given for all metrics. The results show that the influence of $\gamma$ doesn't change by $n$. When $\gamma=0$ (equivalent to cumulative degree predictor), the AUC and $P_n$ are relatively small and $Q_n=0$ for all data sets. This indicates that PA have little efficacy in predicting the future popularity. A small $\gamma$ (a relatively slow time decay) can significantly increase the prediction performance, especially for $Q_n$. The high value of $Q_n$ indicates that the temporal-based predictor has a great power to identify "new objects" which are not yet popular. A too large $\gamma$ will decrease the performance, and we can get the best performance of TBP for all data sets by changing $\gamma$. We denote $\gamma^*$ as the parameter resulting in the highest $P_n$ value. It is clear that the performance of TBP with parameter $\gamma^*$ is remarkably higher than that of PA (i.e. $\gamma=0$ in TBP).

Table \ref{tab2} shows the performance of TBP and PBP for the three data sets. The parameter for each predictor is selected as the one corresponding to the highest $P_n$ value. From the results, we could find that TBP has a better performance for most evaluation metrics. The best $\lambda$ value for PBP is 0.98 for both Movielens and Netflix data set and 0.93 for Facebook data set, which indicates that the links an object received long time ago has a small influence on its future popularity. Unlike the arbitrarily dividing the time into two segments in PBP, TBP uses an exponential decay function which can future improve the prediction performance.

\begin{figure}[h!]
\centering
\scalebox{0.6}[0.6]{\includegraphics{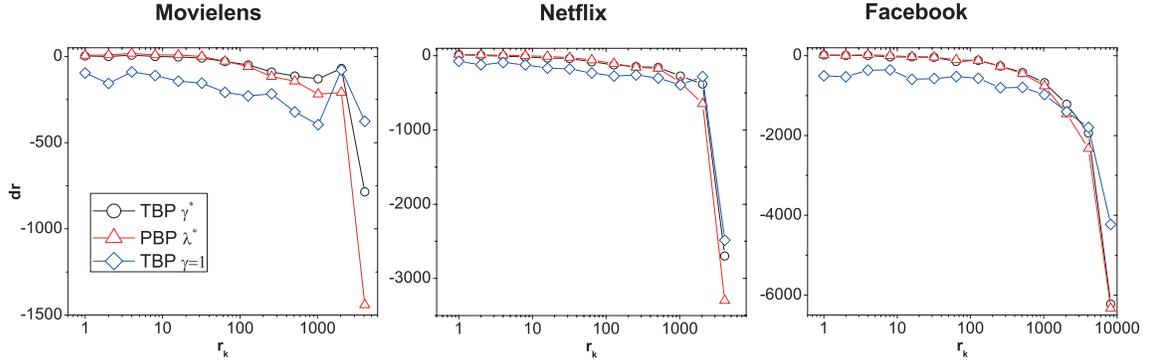} }
\caption{
{\bf The correlation of $dr$ with $r_k$ for top 100 objects in the real future.} Black lines with circles present the result of TBP with the best parameter $\gamma^*$, red lines with triangles present the result of PBP with the best parameter $\lambda^*$, and blue lines with diamonds present the result of TBP with $\gamma=1$. $T_F$ is set as 30 days for all data sets.
}
\label{ana}
\end{figure}

We set a rank change value for each object as $dr=r_f-r_p$, where $r_f$ is the real rank in the near future and $r_p$ is the rank by a predictor. $dr=0$ indicates the prediction rank is echo to the real rank in the near future, and the predictor has perfect performance; $dr<0$ indicates the predictor underestimate the object's popularity. $dr>0$ indicates the predictor overrate the object's popularity. To test how different predictors influence the rank of the top objects in the future, we plot figure \ref{ana} to show the correlation of $dr$ with these objects' degree rank $r_k$ in the testing time under different predictors and parameters. This figure only considers the top 100 objects in the $T_F$ window. The parameters $\lambda$ for PBP are set as the one corresponding to the highest $P_n$ value. The parameters $\gamma^*$ and $\gamma=1$ are also selected for TBP in figure \ref{ana}. These objects are ranked in the top 100 positions in the near future, but their history popularity has a board distribution. Both TBP and PBP with the best parameter can reduce the absolute value of $dr$, which makes these predictors have better performance in predicting future popularity. Compered with PBP, TBP is better at improving the rank of objects that have high $r_k$ but ranked in top place in the future. When $\gamma=1$ in TBP, it is clear that the absolute value of rank difference $d_r$ of the objects that have high $r_k$ becomes smaller than the case with $\gamma^*$, but the absolute value of rank difference $d_r$ of objects with lower $r_k$ becomes larger. That may be due to the fact that parameter $\gamma$ can give less rank score to the objects that are popular in the past, and then improve the rank of the objects that are not popular in the past. The higher the parameter $\gamma$ is, the more obvious this influence is.

\begin{figure}[h!]
\centering
\scalebox{0.7}[0.7]{\includegraphics{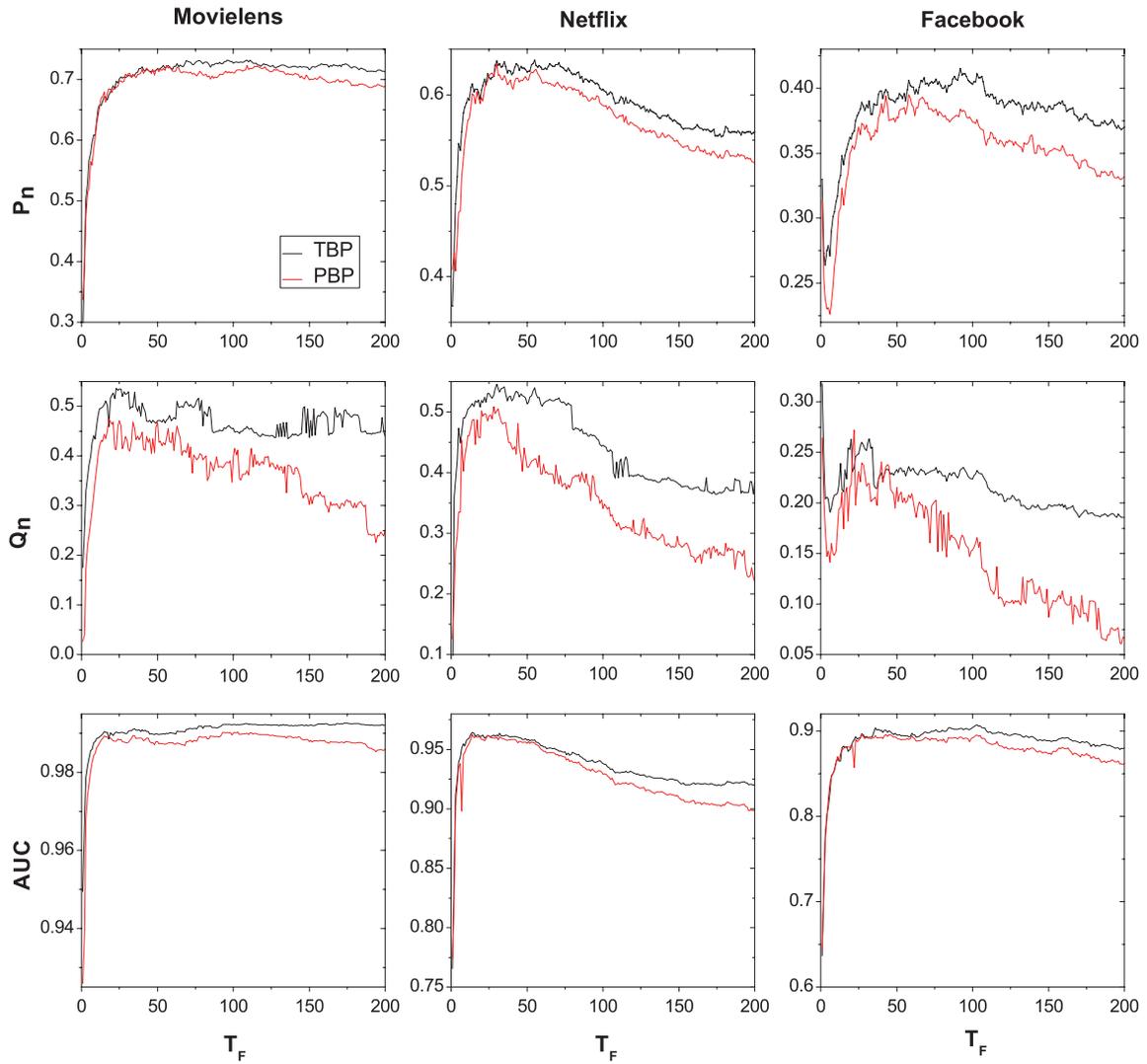} }
\caption{
{\bf The performanc of TBP and PBP as a function of the future time window $T_F$.} TBP is presented by black lines, and PBP is presented by red line. $n$ is set as 100. Time is measured in days. }
\label{comp}
\end{figure}

A small $T_F$ aims to predict objects' popularity in the short term while a large $T_F$ requires to predict the trend in long term. Therefore, we test the performance of the predictors under different $T_F$. Figure \ref{comp} shows the performance of the TBP and PBP as a function of the future time window $T_F$. The parameters corresponding to the highest $P_n$ value for each predictors at each $T_F$ point are selected. For PBP, $T_P$ is set as the same length of $T_F$. Compared with PBP, it is clear that TBP has a better prediction performance for all $T_F$ value. For all data sets, the precision $P_n$, novelty $Q_n$ and AUC of the predictors increase substantially with $T_F$ when $T_F$ is very small. This is because there is a lot of noise when $T_F$ is too small. However, the precision decreases with $T_F$ while $T_F$ becomes larger. This may because the predicted popularity becomes outdated for larger $T_F$ time.

\begin{figure}[h!]
\centering
\scalebox{0.6}[0.6]{\includegraphics{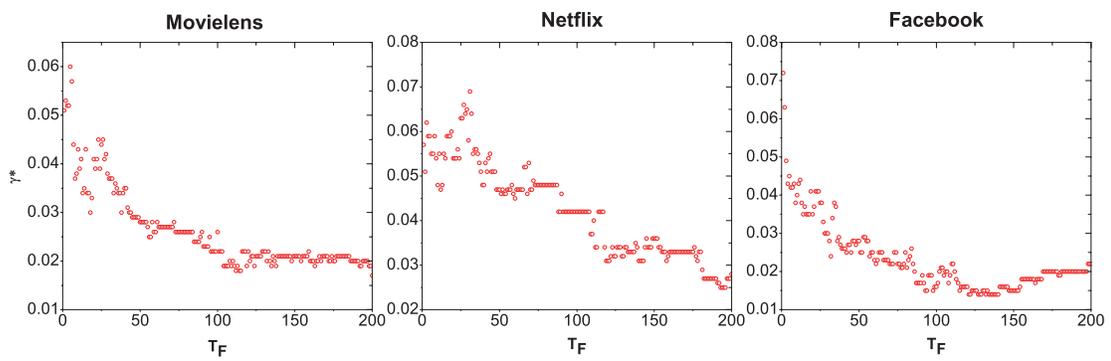} }
\caption{
{\bf The relationship of $\gamma^*$ with $T_F$ for Movielens, Netflix and Facebook data sets.}
}
\label{lambda}
\end{figure}

As we know, $\gamma$ can control the decay speed of the influence of the old links. For different prediction time interval $T_F$, the best parameter $\gamma^*$ may be different. Figure \ref{lambda} shows the relationship of $\gamma^*$ with $T_F$. One could find that, the larger the $T_F$ length, the smaller the $\gamma^*$ value. That means for a shorter $T_F$ prediction, the objects' recent popularity matters more. While for a larger $T_F$ interval prediction, longer historical popularity should be considered.

\section*{Discussion}
To summarize, we proposed a temporal-based predictor (TBP) in this paper, and studied the performance of TBP in trend prediction. The basic idea of TBP is to introduce an exponentially time decay to predict objects' future popularity. We make use of three metrics to evaluate the predictor's performance: $P_n$, $Q_n$ and AUC. We found that the parameter $\gamma$, which controls the speed of the time decay, can give less rank score to the objects that are popular in the past, and accordingly improve the rank of the objects that are not popular in the past. The higher $\gamma$ is, the more obvious of this influence is. Thus, TBP has a higher ability to detect "new entries" that have a lower cumulative popularity but a higher future popularity, and promote these objects to the front of the predicted ranking list. Compared with PBP, TBP has a higher ability to detect the objects that will be popular in the future with different future length $T_F$.

Ranking is one of the most important and fundamental method to solve information over load problem. The study of the popularity dynamic of online information gives us some inspiration to solve the trend prediction problem. This is a very practical issue. In this paper, we studied the links' temporal effects on objects' future popularity. This study is based on the experimental finding that the ability of a node to attract new links vanishes exponentially with time. Besides time, there are a lot of other elements that can influence the dynamic of popularity, such as human dynamics, links heterogeneity, and external influence. Introducing the influence of these elements may future improve the performance of trend predictor. In our future studies, we will focus on improving the performance of the trend predictors with the help of both empirical observations and theoretical analysis.

\section*{Acknowledgments}

\bibliography{bibfile}

\begin{thebibliography}{10}
\providecommand{\url}[1]{\texttt{#1}}
\providecommand{\urlprefix}{URL }
\expandafter\ifx\csname urlstyle\endcsname\relax
  \providecommand{\doi}[1]{doi:\discretionary{}{}{}#1}\else
  \providecommand{\doi}{doi:\discretionary{}{}{}\begingroup
  \urlstyle{rm}\Url}\fi
\providecommand{\bibAnnoteFile}[1]{%
  \IfFileExists{#1}{\begin{quotation}\noindent\textsc{Key:} #1\\
  \textsc{Annotation:}\ \input{#1}\end{quotation}}{}}
\providecommand{\bibAnnote}[2]{%
  \begin{quotation}\noindent\textsc{Key:} #1\\
  \textsc{Annotation:}\ #2\end{quotation}}
\providecommand{\eprint}[2][]{\url{#2}}

\bibitem{simon}
Simon HA (1971) Designing organizations for an information-rich world.
\newblock Computers, communications, and the public interest 72: 37.
\bibAnnoteFile{simon}

\bibitem{news}
Szabo G, Huberman BA (2010) Predicting the popularity of online content.
\newblock Communications of the ACM 53: 80.
\bibAnnoteFile{news}

\bibitem{asur2011trends}
Asur S, Huberman BA, Szabo G, Wang C (2011) Trends in social media: persistence
  and decay.
\newblock In: ICWSM.
\bibAnnoteFile{asur2011trends}

\bibitem{blogs}
Kumar R, Novak J, Raghavan P, Tomkins A (2005) On the bursty evolution of
  blogspace.
\newblock World Wide Web 8: 159.
\bibAnnoteFile{blogs}

\bibitem{youtube}
Crane R, Sornette D (2008) Robust dynamic classes revealed by measuring the
  response function of a social system.
\newblock Proceedings of the National Academy of Sciences 105: 15649.
\bibAnnoteFile{youtube}

\bibitem{popularityyoutube}
Figueiredo F, Fabr\'{i}cio B, Jussara AM (2011) The tube over time:
  characterizing popularity growth of youtube videos.
\newblock In: In Proceedings of the fourth ACM international conference on Web
  search and data mining. ACM, pp. 745-754.
\bibAnnoteFile{popularityyoutube}

\bibitem{bbs}
Aperjis C, Huberman BA, Wu F (2010) Harvesting collective intelligence:
  Temporal behavior in yahoo answers.
\newblock ArXiv:1001.2320.
\bibAnnoteFile{bbs}

\bibitem{product}
Gruhl D, Guha R, Kumar R, Novak J, Tomkins A (2005) The predictive power of
  online chatter.
\newblock In: Proceedings of the eleventh ACM SIGKDD international conference
  on Knowledge discovery in data mining. ACM, pp. 78-87.
\bibAnnoteFile{product}

\bibitem{lerman2010predict}
Lerman K, Tad H (2010) Using a model of social dynamics to predict popularity
  of news.
\newblock In: In Proceedings of the 19th international conference on World wide
  web. ACM, pp. 621-630.
\bibAnnoteFile{lerman2010predict}

\bibitem{BA}
Barab\'{a}si AL, Albert R (1999) Emergence of scaling in random networks.
\newblock Science 286: 509.
\bibAnnoteFile{BA}

\bibitem{Nopa}
Newman MEJ (2009) The first-mover advantage in scientific publication.
\newblock Europhysics Letters 86: 68001.
\bibAnnoteFile{Nopa}

\bibitem{Nopaim}
Newman MEJ (2014) Prediction of highly cited papers.
\newblock Europhysics Letters 105: 28002.
\bibAnnoteFile{Nopaim}

\bibitem{matusPRL}
Medo M, Cimini C, Gualdi S (2011) Temporal effects in the growth of networks.
\newblock Physical review letters 101: 238701.
\bibAnnoteFile{matusPRL}

\bibitem{infoAccess}
Dezs\"{o} Z, Almaas E, Luk\'{a}cs A, R\'{a}cz B, Szakad\'{a}t I, et~al. (2006)
  Dynamics of information access on the web.
\newblock Physical Review E 73: 066132.
\bibAnnoteFile{infoAccess}

\bibitem{busrtM}
Ratkiewicz J, Fortunato S, Flammini A, Menczer F, Vespignani A (2010)
  Characterizing and modeling the dynamics of online popularity.
\newblock Physical review letters 105: 158701.
\bibAnnoteFile{busrtM}

\bibitem{busrtWeibo}
Yan Q, Wu L (2012) Impact of bursty human activity patterns on the popularity
  of online content.
\newblock Discrete Dynamics in Nature and Society .
\bibAnnoteFile{busrtWeibo}

\bibitem{timeSeriesShap}
Yang J, Leskovec J (2011) Patterns of temporal variation in online media.
\newblock In: Proceedings of the fourth ACM international conference on Web
  search and data mining. ACM, pp. 177-186.
\bibAnnoteFile{timeSeriesShap}

\bibitem{asur2012}
Roja B, Asur S, Huberman BA (2012) The pulse of news in social media:
  Forecasting popularity.
\newblock In ICWSM .
\bibAnnoteFile{asur2012}

\bibitem{diggp}
Jamali S, Rangwala H (2009) Digging digg: Comment mining, popularity
  prediction, and social network analysis.
\newblock In: In Web Information Systems and Mining. International Conference
  on. IEEE, pp. 32-38.
\bibAnnoteFile{diggp}

\bibitem{zenganTp}
Zeng A, Gualdi S, Medo M, Zhang YC (2013) Trend prediction in temporal
  bipartite networks: the case of movielens, netflix, and digg.
\newblock Advances in Complex Systems 16: 04n05.
\bibAnnoteFile{zenganTp}

\bibitem{wang2013quantifying}
Wang D, Song C, Barab{\'a}si AL (2013) Quantifying long-term scientific impact.
\newblock Science 342: 127--132.
\bibAnnoteFile{wang2013quantifying}

\bibitem{ZhuDecay}
Zhu H, Wang X, Zhu JY (2003) Effect of aging on network structure.
\newblock Physical Review E 68: 056121.
\bibAnnoteFile{ZhuDecay}

\bibitem{konect:2014:facebook-wosn-wall}
Facebook wall posts network dataset - {KONECT}.
\newblock Available:
  \url{http://konect.uni-koblenz.de/networks/facebook-wosn-wall}.
\newblock Accessed 20 September 2014.
\bibAnnoteFile{konect:2014:facebook-wosn-wall}

\bibitem{viswanath09}
Viswanath B, Mislove A, Cha M, Gummadi KP (2009) On the {Evolution} of {User}
  {Interaction} in {Facebook}.
\newblock In: Proc. Workshop on Online Social Networks. pp. 37-42.
\bibAnnoteFile{viswanath09}

\bibitem{konect}
Kunegis J (2013) {KONECT} - {The} {Koblenz} {Network} {Collection}.
\newblock In: Proc. Int. Web Observatory Workshop. pp. 1343-1350.
\bibAnnoteFile{konect}

\bibitem{konect-network}
The koblenz network collection.
\newblock Available: \url{http://konect.uni-koblenz.de/}.
\newblock Accessed 30 November 2014.
\bibAnnoteFile{konect-network}

\bibitem{AUC}
Hanley JA, McNeil BJ (1982) The meaning and use of the area under a receiver
  operating characteristic (roc) curve.
\newblock Radiology 143: 29.
\bibAnnoteFile{AUC}

\bibitem{Precision}
Herlocker JL, Konstan JA, Terveen LG, Riedl JT (2004) Evaluating collaborative
  filtering recommender systems.
\newblock ACM Transactions on Information Systems (TOIS) 22: 5-53.
\bibAnnoteFile{Precision}

\end{thebibliography}

\section*{Figure Legends}
%

\section*{Tables}
%
%
%

\begin{table}[!ht]
\center
\footnotesize
\caption{ Basic statistical features of the data sets. }
\label{tab1}
\setlength\tabcolsep{20pt}
\begin{tabular}{ccccc}

\hline
Data set& Users & Objects & Links & Period \\
\hline
\centering
Movilens & 5000 & 7533 & $8.6\times 10^5$ & 1st Jan 2002 - 1st Jan 2005\\				
Netflix & 4960 & 16599 & $1.2 \times 10^6$ & 1st Jan 2000 - 31st Dec 2005\\
Facebook & 40981 & 38143  & $8.6 \times 10^5$ & 14th Sep 2004 - 22nd Jan 2009\\
\hline
\end{tabular}
\end{table}

\begin{table}[!ht]
\center
\footnotesize
\caption{The prediction performance of temporal-based predictor (TBP) and popularity-based predictor (PBP) for Movielens, Netflix and Facebook data sets. The parameter for each predictor is set as the one corresponding to the highest $P_n$ value. $n$ is set as 100.}
\label{tab2}
\setlength\tabcolsep{18pt}
\begin{tabular}{ccccccc}

\hline
Data set & Predictor & Parameter  & AUC & $P_n$ & $Q_n$ \\
\hline
\centering
\multirow{2}*{Movielens} & PBP & 0.98 & 0.988	& 0.706 & 0.432 \\
 & TBP & 0.06 & 0.990 & 0.705 & 0.526	\\
\hline

\hline
\centering
\multirow{2}*{Netflix} & PBP & 0.98	 &	0.960 & 0.634 &	0.502	\\
& TBP & 0.06	& 0.962 & 0.637 & 0.545	\\
\hline

\hline
\centering
\multirow{2}*{Facebook} & PBP & 0.93 &	0.893 & 0.372 &	0.217	\\	
& TBP & 0.03 & 0.894 & 0.387 & 0.252	\\
\hline
\end{tabular}
\end{table}

\end{document}